\documentclass[showpacs,floatfix,superscriptaddress,showpacs,preprint,amssymb,amsfonts,prb,aps]{revtex4}
\usepackage{longtable,graphicx,epsfig,dcolumn}

\begin{document}
\bibliographystyle{revtex}
\title{Resonant X-ray Reflectivity from the Liquid Bi-In Surface}

\author{Elaine~DiMasi}
\affiliation{Department of Physics, Brookhaven National
Laboratory, Upton NY 11973-5000}

\author{Holger~Tostmann}
\affiliation{Division of Applied Sciences and Department of
Physics, Harvard University, Cambridge MA 02138}

\author{Oleg~G.~Shpyrko}
\affiliation{Division of Applied Sciences and Department of
Physics, Harvard University, Cambridge MA 02138}

\author{Patrick~Huber}
\affiliation{Division of Applied Sciences and Department of
Physics, Harvard University, Cambridge MA 02138}

\author{Benjamin~M.~Ocko}
\affiliation{Department of Physics, Brookhaven National
Laboratory, Upton NY 11973-5000}

\author{Peter~S.~Pershan}
\affiliation{Division of Applied Sciences and Department of
Physics, Harvard University, Cambridge MA 02138}

\author{Moshe~Deutsch}
\affiliation{Department of Physics, Bar-Ilan University, Ramat-Gan
52100, Israel}

\author{Lonny~E.~Berman}
\affiliation{National Synchrotron Light Source,
    Brookhaven National Laboratory, Upton NY 11973}

\begin{abstract}
Resonant x-ray reflectivity measurements from the surface of
liquid Bi$_{22}$In$_{78}$
find only a modest surface Bi enhancement, with 35~at\% Bi
in the first atomic layer.
This is in contrast to the Gibbs adsorption in all liquid alloys
studied to date, which show surface segregation of
a complete monolayer of the low surface tension component.
This suggests that surface adsorption in Bi-In is dominated by
attractive interactions that increase the number of Bi-In neighbors
at the surface.
These are the first measurements in which resonant x-ray scattering has been
used to quantify compositional changes induced at a liquid alloy surface.
\end{abstract}


\maketitle

\section{Introduction}

Current treatments of the thermodynamics of surface phenomena in solutions
rely heavily on the original works by Gibbs in 1878, and
one of the most familiar corollaries is the Gibbs adsorption rule.
In its simplest invocation, the Gibbs rule states that in a binary liquid,
the species having the lower surface tension will segregate preferentially
at the surface.
This apparent simplicity is deceptive:
a survey of the literature reveals a hundred years'
debate over the application of the Gibbs adsorption
    rule,\cite{cahn58} 
not to mention its extension to multicomponent
    systems\cite{widom79}
and crystalline
    surfaces,\cite{rusanov96} 
and its connection to atomistic
    models.\cite{speiser87}

Experimental investigations of the validity of the Gibbs rule encompass
measurements of adsorption
    isotherms,\cite{adamson67}
    surface tension,\cite{bhatia78} 
and surface
    composition\cite{hardy82} 
in a variety of systems.
Unfortunately, many of the liquids studied are too complicated
for the simplest formulations of the Gibbs rule.
Liquid metal alloys are in many ways ideal for such studies.
Miscible alloys exist which behave as ideal liquids, while in other systems
strongly attractive or repulsive heteroatomic interactions can be studied.
Perhaps an even more important advantage of liquid metals is that the
compositionally inhomogeneous region at the surface is known in some cases
to be confined to an atomic layer.
This is commonly assumed in calculations of Gibbs adsorption that
take a model of a physical surface as their starting point.

For example, x-ray reflectivity, ion scattering, and Auger electron
spectroscopy measurements of liquid Ga$_{84}$In$_{16}$
found a 94\% In surface monolayer, as expected
given this alloy's positive heat of
    mixing.\cite{regan97,dumke83}
Subsequent layers have the bulk composition.
Similar studies of dilute liquid Bi-Ga ($<0.2$~at\% Bi)
likewise found surface segregation of a pure Bi
    monolayer.\cite{lei96}
Even when the repulsive
interactions between Ga and Bi cause more Bi-rich alloys to undergo
additional phase separation above 220$^{\circ}$C,
where a 65~\AA\ thick inhomogeneous Bi-rich region forms, the pure Bi
surface monolayer
    persists.\cite{srn99,gabi00}
In Ga-Bi, then, repulsive heteroatomic interactions substantially
change the surface composition profile, but do not defeat the Gibbs
adsorption.

The effect of attractive heteroatomic interactions remains an open question.
In alloys such as Bi-In, attractive forces
between the two species produce a number of compositionally ordered phases
in the bulk solid.
It is therefore conceivable that Bi-In pairing may exist at the liquid
surface, and compete with surface segregation.
Our recent temperature dependent x-ray reflectivity measurements
of liquid Bi-In alloys having 22, 33, and 50~at\% Bi
revealed structural features not found in
elemental metals or in the Ga alloys discussed
    above.\cite{ben00}
As we will show, those data were suggestive of Bi-In pair formation along
the surface-normal direction.  However, since the technique did not measure
the Bi surface concentration directly, other interpretations of the data
were also possible.

 \section{Experimental Setup}
A complete characterization of surface composition requires both elemental
specificity and {\AA}-scale structural resolution along the
surface-normal direction, which is difficult to achieve experimentally.
Auger electron spectroscopy, which satisfies the first of these requirements,
is hampered by contributions from the bulk liquid.\cite{hardy82}
X-ray reflectivity by contrast is a surface-sensitive probe.
In the kinematic
    limit\cite{born}
the reflected intensity, measured as a function of momentum transfer
$q_z$ normal to the surface, is
proportional to the Fresnel reflectivity $R_F$ of a homogeneous
    surface:\cite{refl}
\begin{equation}
\label{biin_eq:refl}
    R(q_z) = R_F \left| (1/\rho_\infty)
        \int_{-\infty}^\infty
        (\partial \rho_{\mbox{\scriptsize eff}} / \partial z)
        \exp(i q_z z) \, dz \right|^2 \: .
\end{equation}
Here $\rho_{\mbox{\scriptsize eff}}$ represents an effective electron
scattering amplitude that combines the electron density profile
with the scattering form factor, and
$\rho_\infty$ is the density of the bulk. The electron density
variations that produce modulations in the reflectivity may result
from changes in either the composition or the mass density. Thus,
inference of surface composition from the measured reflectivity is
sometimes ambiguous.

 \section{Resonant X-ray Scattering}
This disadvantage can be overcome with the application of resonant x-ray
scattering. The effective electron density of a scattering atom depends on
the scattering form factor
$f(q)+f^{\prime}(q,E)  \approx Z+f^{\prime}(E)$.
When the x-ray energy is tuned to an absorption edge of a scattering atom,
the magnitude of $f^{\prime}$ becomes appreciable, producing changes in
contrast between unlike
    atoms.\cite{resonant}
With one exception,\cite{hgau00}
resonant x-ray scattering measurements reported in the past have been
confined to studies of solids and bulk liquids,
due to the difficulty of the experiment.
The present report is the
first to find compositional changes induced at a liquid surface.

\begin{figure}[tbp]
\centering
\includegraphics[angle=90,width=1.0\columnwidth]{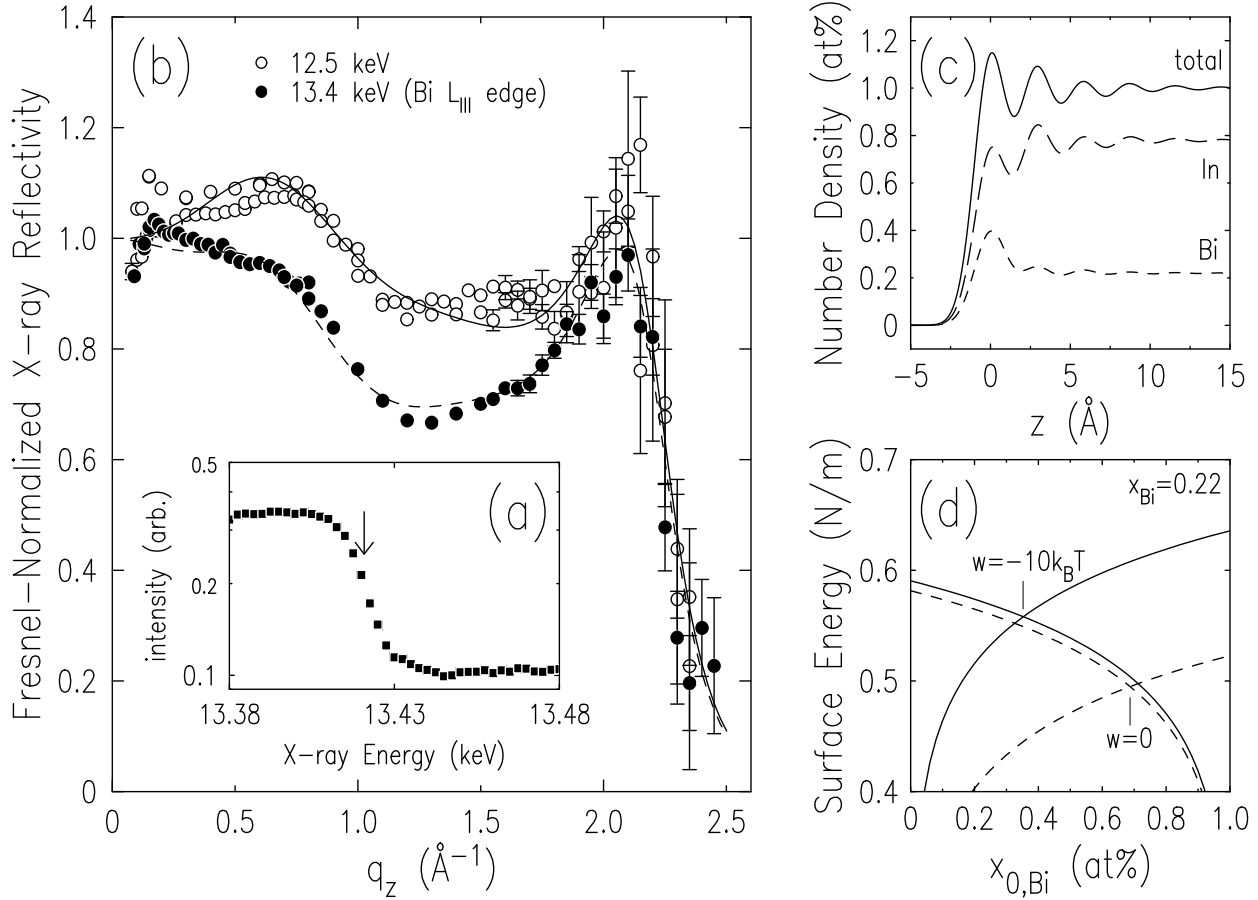}
\caption{ (a) Energy scan in transmission through Bi foil. (b)
Fresnel-normalized x-ray reflectivity of Bi$_{22}$In$_{78}$.
$\circ$: 12.5 keV (two independent measurements); $\bullet$: 13.4
keV; (---): 12.5 keV fit; (- - -) 13.4 keV fit. (c) Best fit real
space number density profile relative to bulk atomic percent.
(---): Total number density; (-- --): In density; (- - -): Bi
density. (d) Surface energy as a function of surface Bi
concentration $x_{0,\mbox{\scriptsize Bi}}$ for the bulk
concentration $x_{\mbox{\scriptsize Bi}} = 0.22$, according to
Eq.~\ref{biin_eq:refl}. (- - -): $w=0$ .  (---): $ w = - 10 k_B
T$. } \label{biin_fig:1}
\end{figure}

 \section{Sample Preparation}
The molten Bi$_{22}$In$_{78}$ sample was maintained at
$T= 101^\circ$C and
$P = 5 \times 10^{-10}$~Torr within an ultra high vacuum chamber,
and periodically sputter cleaned with Ar$^+$ ions.
Reflectivity measurements were performed at beamline X25
at the National Synchrotron Light Source.
The spectrometer has been described
    previously,\cite{indium}
except that here a double Si(111) crystal monochromator
was used to provide an energy resolution of 9~eV.  For the present
work, we compare reflectivity measured at 12.5 keV to measurements made at
the Bi $L_{III}$ edge at 13.421 keV. The energy was calibrated by
transmission through a Bi foil, shown in
    Figure~\ref{biin_fig:1}(a).
At the inflection point indicated by the arrow,
$f^{\prime}_{\mbox{\scriptsize Bi}}$ has its largest magnitude of $-24.7$
electrons.
Uncertainties in $f^\prime_{\mbox{\scriptsize Bi}}$
may arise from inaccuracies in the calculation,
incorrect establishment of the incident energy, and the energy resolution.
To account for these possibilities, the analysis was performed for deviations
in $f^\prime_{\mbox{\scriptsize Bi}}$ of about
20\% (i.e., $f^\prime_{\mbox{\scriptsize Bi}}=-19.7$ and $-29.7$ electrons).
The results were incorporated into the error ranges tabulated below.

 \section{X-Ray Reflectivity Data}
Reflectivity data at both energies are shown in
    Figure~\ref{biin_fig:1}(b)
(symbols).  These results were found to be both reproducible in energy
and stable over time by measuring two 12~keV data sets, prior to and
following the 13.4~keV measurements.  The two 12~keV data sets are shown
together as open circles in
    Figure~\ref{biin_fig:1}(b).
The interference peak at $q_z = 2.1$~\AA$^{-1}$
is due to stratification of the atoms in planes parallel to the surface,
a well established feature of liquid
    metals.\cite{rice}
For $q_z < 1.8$~\AA$^{-1}$, the data exhibit a modulation
indicative of a structural periodicity roughly twice that of the surface
layering.  This feature is consistent with Bi-In
dimers oriented along the surface normal, which could also give rise to
alternating Bi and In layers (bilayers) at the surface.
We also find that the low-$q_z$ reflectivity is strongly decreased when
measured at the Bi $L_{III}$ edge.
The reflectivity decrease itself  varies smoothly with $q_z$
and does not exhibit a bilayer-type modulation,
instead suggesting that the surface Bi
concentration is larger than that of the bulk.

 \section{Electron Density Model}
To investigate these possibilities,
we calculate the reflectivity of a model density profile according to
    Eq.~\ref{biin_eq:refl},
which is refined simultaneously against the data taken at both energies.
Resonant effects are included by combining the model structure and the
scattering amplitude into an effective electron density profile
$\rho_{\mbox{\scriptsize eff}}(z)$.
The energy dependence enters the analysis through the
Bi concentration defined in
$\rho_{\mbox{\scriptsize eff}}(z)$,
and also through the Fresnel
reflectivity $R_F$, which is a function of the energy dependent mass
absorption coefficient $\mu^{-1}$ and the effective bulk electron density
$\rho_\infty \propto (Z-f^{\prime})$
that defines the critical angle $q_c$.
    Table~\ref{biin_table:pars}
shows the values of these quantities used in our models.
Reflectivity data acquired at each
energy were normalized to the appropriate energy dependent Fresnel function.

\begin{table}[tcp]

\begin{tabular}{ccccccccc}
    & E  & $\mu / \rho_m $$^a$ & $\rho_m $$^b$
    & $\mu ^{-1} $ & $f^{\prime \; c}$
    & $Z-f^{\prime}$ & $\rho _\infty$  & $q_c$  \\

    & (keV) & (cm$^2$/g) &  (g/cm$^3$) & ($\mu$m) & &
    & ($e^- /$\AA$^3$) & (\AA$^{-1}$) \\ \hline
In: & 12.5  & 74.1 & 7.0 & 19.3 & -0.2 & 48.8 & 1.788 & --- \\
& 13.4  & 60.5 &  & 23.6 &  &  &  & --- \\ \hline
Bi: & 12.5  & 75.2 & 10.0 & 13.3 & -7.3 & 75.7 & 2.180 & --- \\
& 13.4  & 155.0 &  & 6.45 & -24.7 & 58.3 & 1.679 & --- \\
\hline Bi$_{22}$In$_{78}$: & 12.5  & 74.5 & 7.66 & 17.5 & --- &
--- & 1.874 &
0.05154 \\
& 13.4  & 92.6 &  & 14.1 & --- & --- & 1.764 & 0.05000
\end{tabular}
\\

$^a$C.~H.~MacGillavry and G.~D.~Rieck, eds., {\em International
Tables for X-ray Crystallography} Vol.\ III, Kynoch Press,
Birmingham, England (1962). $^b$R.~C.~Weast, ed., {\em CRC
Handbook of Chemistry and Physics\/} (see Ref.\ 20).
$^c$B.~L.~Henke, E.~M.~Gullikson, and J.~C.~Davis, Atomic Data and
Nuclear Tables {\bf 54} (1993) 181. \caption{ Parameters used to
calculate energy dependent x-ray reflectivity. }
\label{biin_table:pars}
\end{table}

Following past
    practice,\cite{indium}
our model incorporates layers of atoms having a Gaussian distribution of
displacements
from idealized positions $nd$ along the surface normal direction:
\begin{equation}
    \rho_{\mbox{\scriptsize eff}} (z)
        =\rho _{\infty } \sum_{n=0}^{\infty }F_{n}
    \frac{d}{\sigma _{n}\sqrt{ 2\pi }}\exp
    \left[ -(z-nd)^{2}/\sigma _{n}^{2}\right] .
\label{biin_eq:layers}
\end{equation}
The roughness $\sigma _n$ arises from both static and dynamic contributions:
\begin{equation}
    \sigma _{n}^{2}=n\overline{\sigma }^{2}+\sigma _{0}^{2}+
    \frac{k_{B}T}{2 \pi \gamma }\ln
    \left( \frac{q_{\mbox{\scriptsize max}}}{q_{\mbox{\scriptsize res}}}
    \right) .
\label{biin_eq:sigma}
\end{equation}
Here $\overline{\sigma }$ and $\sigma _{0}$ are related to
the surface layering coherence length and the amplitude of density
oscillations at the surface. The last term accounts for height
fluctuations produced by capillary waves, and depends on the
temperature $T=101^{\circ }$C,
    the surface tension $\gamma =0.50$~N/m,\cite{gamma}
and wavevector cutoffs $q_{\mbox{\scriptsize max}}=0.99$~\AA $^{-1}$ and
$q_{\mbox{\scriptsize res}} \sim 0.024$~\AA $^{-1}$
(a slowly varying function of $q_z$),
as detailed
    elsewhere.\cite{indium}

The scattering amplitude of each layer depends on the form factor and the
effective electron density in each layer relative to the bulk,
dependent on energy and Bi concentration:
\begin{equation}
    F_{n}=w_{n}\times \frac{x_{n,\mbox{\scriptsize  Bi}}
    \left[ f_{ \mbox{\scriptsize  Bi}}(q_z)+
    f_{\mbox{\scriptsize  Bi}}^{\prime }(E)\right]
    +(1-x_{n,\mbox{\scriptsize Bi}})
    \left[ f_{\mbox{\scriptsize  In}}(q_z)+
    f_{\mbox{\scriptsize  In}}^{\prime }\right] }
    {x_{\mbox{\scriptsize  Bi}}\left[ Z_{\mbox{\scriptsize  Bi}}+
    f_{\mbox{\scriptsize  Bi}}^{\prime }(E)\right]
    +(1-x_{\mbox{\scriptsize Bi}})\left[ Z_{\mbox{\scriptsize  In}}+
    f_{\mbox{\scriptsize  In}}^{\prime }\right] }\times \frac{x_{n,
    \mbox{\scriptsize  Bi}}\rho _{\infty ,\mbox{\scriptsize  Bi}}+
    (1-x_{n,\mbox{\scriptsize Bi}})\rho _{\infty ,\mbox{\scriptsize  In}}}
    {\rho _{\infty,\mbox{\scriptsize  bulk}}} \: .
\label{biin_eq:prefix}
\end{equation}
For the first few layers ($n$=0,~1,~2), the weight $w_{n}$ may
differ from unity and the Bi fraction $x_{n,\mbox{\scriptsize  Bi}}$
can vary from the bulk value $x_{\mbox{\scriptsize  Bi}}=0.22$.

\begin{figure}[tbp]
\centering
\includegraphics[angle=90,width=1.0\columnwidth]{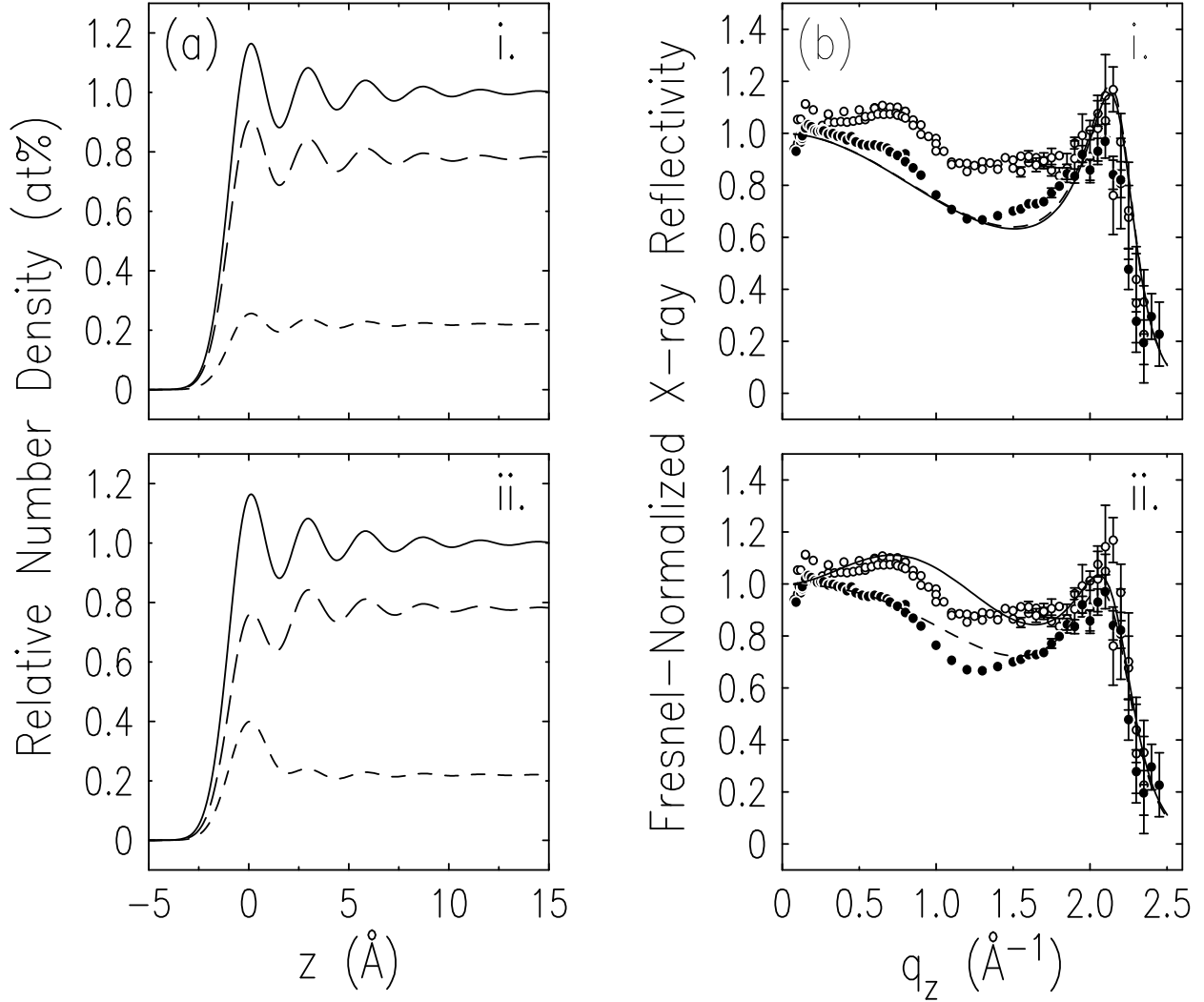}
\caption{ (a) Model surface-normal number density profiles,
relative to bulk atomic percent. (---): total number density; (--
--): In density; (- - -): Bi density. (b) Calculated
Fresnel-normalized reflectivity curves. (---): 12.5 keV; (- - -):
13.4 keV. i: Surface layering with uniform composition
($x_{\mbox{\scriptsize Bi}}=0.22$). ii: Surface layering, with Bi
enhancement in the first atomic layer ($x_{0,\mbox{\scriptsize
Bi}}=0.35$). } \label{biin_fig:2}
\end{figure}

 \section{Data Fitting Analysis}
We now describe the ingredients that are required to fit the data with
this model.  The interference peak at $q_z = 2.1$~\AA$^{-1}$
can be reproduced by a simple layered profile in which
$x_{n,\mbox{\scriptsize Bi}} = x_{\mbox{\scriptsize Bi}} $ for all $n$.
The Bi and In number densities for such a model are shown in
    Figure~\ref{biin_fig:2}(a)i,
along with their sum, the total atomic fraction relative to the bulk.
The corresponding reflectivity
curves calculated for both x-ray energies are compared to the
experimental data in
    Figure~\ref{biin_fig:2}(b)i.
Since the Bi concentration is uniform, the
energy dependence is so slight that the curves overlap  almost completely
on the scale of the figure.
Turning again to the data, the reduced intensity in the region
$q_z < 1.8$~\AA$^{-1}$ when measured at the Bi L$_{III}$
edge indicates a reduction in the scattering amplitude at the surface.
This implies that the Bi concentration is enhanced there.
Increasing the Bi fraction from 22~at\% to
$\approx 35$~at\% in the first layer ($n=0$) produces an appropriate energy
dependence
    (Figs.~\ref{biin_fig:2}(a)ii and \ref{biin_fig:2}(b)ii).

Although at this point the frequency of the low-$q_z$ modulation is not
well described, the fit is considerably improved
by allowing the Bi fraction and total number densities to vary for the
first three surface layers ($n$=0,1,2).
We find that the Bi fraction for $n$=1,2 is essentially equal to the bulk
value of 22~at\%, while the total number densities for $n$=0,1,2 have
values of 0.98, 1.01, and 0.98, respectively
    (Figs.~\ref{biin_fig:1}(b),(c)).
Thus, the detailed shape of the low-$q_z$ modulation is modelled
by a very slight density wave of about 2\% affecting the
amplitudes of the first few surface layers. Increasing the number
of model parameters to allow for shifts in positions and widths of
the surface layers resulted in marginally better fits, but at the
expense of high frequency Fourier components appearing as small
wiggles in the calculated reflectivity.  Variations in these extra
fit parameters are extremely slight, and we doubt whether they
have any physical basis. Parameters for all models are shown in
    Table~\ref{biin_table:2}.

\begin{table}[tbp]
\begin{tabular}{cccccccccc}
Figure & $d$ & $\overline{\sigma}$ & $\sigma_0$ & $w_0$ &
$x_{0,\mbox{\scriptsize Bi}}$ & $w_1$ & $x_{1,\mbox{\scriptsize
Bi}}$ & $w_2$ &
$x_{2,\mbox{\scriptsize Bi}}$ \\
\hline
1 & 2.81 & 0.54 & 0.64 & 0.98 & 0.35 & 1.01 & 0.22 & 0.98 & 0.23 \\
2i & 2.81 & 0.48 & 0.64 & 1.0 & 0.22 & 1.0 & 0.22 & 1.0 & 0.22 \\
2ii & 2.81 & 0.54 & 0.64 & 1.0 & 0.35 & 1.0 & 0.22 & 1.0 & 0.22
\end{tabular}
\caption{Fit parameters for model profiles, identified by the
figure in which they appear. The length scales $d$,
$\overline{\protect\sigma}$, and $\protect\sigma_0$ are in units
of \protect\AA .} \label{biin_table:2}
\end{table}

This analysis demonstrates that
to model the essential features of our data, there is no need to invoke
long-range compositional ordering on a second length scale in the
surface-normal
direction, which we had suggested based on the previous non-resonant
reflectivity
    measurements.\cite{srn99,ben00}
Still, we thought it important to
investigate additional, specific models based on Bi-In pairs oriented
along the surface normal.
To test for alternating Bi-rich and In-rich layers, we attempted fits
in which we forced the Bi and In compositions to
be substantially different from the results shown above. We also described
Bi-In pairing by allowing the positions, but not the densities, of the
surface layers to vary.
None of these profiles successfully described the data.

 \section{Summary}
Our principal finding is the Bi enrichment of 35~at\% in the surface
layer, compared to the bulk value of 22~at\%.  This is considerably less
Bi than would be expected in the absence of attractive Bi-In interactions,
which can be estimated from the surface free energy
    requirement:\cite{gugg}
\begin{equation}
\label{biin_eq:gugg}
    \gamma_{\mbox{\scriptsize In}} + \frac{k_BT}{a}
    \ln \left( \frac {1-x_{0,\mbox{\scriptsize Bi}}}
        {1-x_{\mbox{\scriptsize Bi}}} \right)
    - \frac{1}{4} (x_{\mbox{\scriptsize Bi}})^2 \frac{w}{a}
    = \gamma_{\mbox{\scriptsize Bi}} + \frac{k_BT}{a}
    \ln \left( \frac {x_{0,\mbox{\scriptsize Bi}}}
        {x_{\mbox{\scriptsize Bi}}} \right)
    - \frac{1}{4} (1-x_{\mbox{\scriptsize Bi}})^2 \frac{w}{a} \: .
\end{equation}
The quantity $w$ is the excess interaction energy of Bi-In pairs over
the average of the Bi-Bi and In-In interaction energies; for an ideal
mixture, $w=0$.
This analysis assumes that the inhomogeneous region is confined to a
single atomic layer, the atoms are close-packed and take up an area $a$,
and $w$ is small.
Extrapolating the measured surface tensions to
    100$^\circ$C,\cite{gamma}
$\gamma_{\mbox{\scriptsize In}} = 0.56$~N/m and
$\gamma_{\mbox{\scriptsize Bi}} = 0.41$~N/m.
Using the Bi atomic size, $a =  \pi(3.34/2)^2$~\AA$^2$
    (for In, the atomic diameter is 3.14~\AA ),\cite{iida}
and the bulk composition $x_{\mbox{\scriptsize Bi}} = 0.22$.
The equilibrium surface composition $x_{0,\mbox{\scriptsize Bi}}$ is
found from the intersection of plots of both sides of
    Eq.~\ref{biin_eq:gugg}.
For $w=0$, this analysis predicts a surface segregation of 69~at\% Bi
    (Figure~\ref{biin_fig:2}(d), dashed lines).
To reproduce our experimental finding that
$x_{0,\mbox{\scriptsize Bi}} = 35$~at\%,
$w$ must be negative, with a magnitude of $\sim 10 k_B T$
    (Figure~\ref{biin_fig:2}(d), solid lines).
Although this large value of $w$ is most likely outside the range of
validity of
    Eq.~\ref{biin_eq:gugg},
the analysis certainly illustrates the qualitative effect of attractive
heteroatomic interactions on the surface composition.
In this Bi-In alloy, pairing does in fact defeat Gibbs adsorption in the
sense that the surface energy is optimized not by segregating a large
fraction of Bi,
but by forming larger numbers of Bi-In neighbors in the surface layer.
Exactly how this balance plays out in Bi-In alloys with the stoichiometric
bulk compositions BiIn and BiIn$_2$ remains to be seen.


\bibliographystyle{unsrt}

\begin{thebibliography}{1}
\bibitem{born} The kinematic limit, or Born approximation, is valid where
$R(q_z) << 1$.  In this work, data are fit only for $q_z > 0.2$, where
$R(q_z) < 3 \times 10^{-4}$, and the structure is determined principally
by data with $q_z > 0.4$, where $R(q_z)$ is in the range
$10^{-5}$--$10^{-9}$.
\end{thebibliography}

\end{document}